\documentstyle[aps,epsf,twocolumn]{revtex}
\begin{document}
\draft

\title{\Large \bf Short-Range Ising Spin Glass: Multifractal Properties.}

\author{\bf E.\ Nogueira Jr.}
\address{Departamento de F\'\i sica,\\
Universidade Federal do Rio Grande do Norte,\\
CP 1641, 59072-970, Natal, RN, Brazil.}

\author{\bf S.\ Coutinho}
\address{Laborat\'orio de F\'\i sica Te\'orica e Computacional\\
Universidade Federal de Pernambuco,\\
50670-901, Recife, PE, Brazil.}

\author{ \bf F.\ D.\ Nobre}
\address{Departamento de F\'\i sica,\\
Universidade Federal do Rio Grande do Norte,\\
CP 1641, 59072-970, Natal, RN, Brazil.}

\author{\bf E.\ M.\ F.\ Curado\cite{iccmp}}
\address{Centro Brasileiro de Pesquisas F\'\i sicas,\\
Rua Xavier Siguad 150, \\
222290-180, Rio de Janeiro, RJ, Brazil.}

\author{ \bf J.\ R.\ L.\ de Almeida}
\address{Departamento de F\'\i sica, \\
Universidade Federal de Pernambuco,\\
50670-901, Recife, PE, Brazil.}

\date{Accepted to Phys. Rev. E /01 March 97}

\maketitle

\begin{abstract}
The multifractal properties of the Edwards-Anderson order parameter of the
short-range Ising spin glass model on $d=3$ diamond hierarchical lattices is
studied via an exact recursion procedure. The profiles of the local order
parameter are calculated and analysed within a range of temperatures close
to the critical point with four symmetric distributions of the coupling
constants (Gaussian, Bimodal, Uniform and Exponential). Unlike the pure
case, the multifractal analysis of these profiles reveals that a large
spectrum of the $\alpha $-H\"older exponent is required to describe the
singularities of the measure defined by the normalized local order
parameter, at and below the critical point. Minor changes in these spectra
are observed for distinct initial distributions of coupling constants,
suggesting an {\em universal} spectra behavior. For temperatures slightly
above $T_{c}$, a dramatic change in the $F(\alpha )$ function is found,
signalizing the transition.
\end{abstract}

\pacs{05.50.+q, 75.10.Nr and 64.60.A}

\section{Introduction}

The understanding of the nature of the spin-glass (SG) condensed phase in
real systems has been challenging many authors\cite{Review}, since the
scenario emerging from Parisi's mean-field solution \cite{Parisi} of the
Sherrington-Kirkpatrick (SK) \cite{SK} model came out. Some raised
conclusions, like the structure of the free-energy barriers corresponding to
many distinct phases below $T_{c}$ (pure states) \cite{Vertechi},
arranged in an ultrametric structure and the existence of a critical
ordering field for the condensed phase\cite{Almeida}, generated
controversies that remain not satisfactorily elucidated. In particular, the
domain-wall phenomenological scaling approach ({\em droplet model}) dismiss
the SK model as appropriated for the description of short-range Ising spin
glasses in low dimensions and does not share the same conclusions \cite
{Bray82,Fisher}. On the other hand, recent works based on numerical
simulations presented results indicating that short-range models should
exhibit the same qualitative features appearing in the SK model \cite
{Georges}. It is worth to mention that many efforts have been devoted to
investigate exactly-solvable short-range SG models as an attempt to describe
real spin glasses, where certain aspects of the system, e.g., the
correlation length, the sensibility to the boundary conditions, and
finite-size effects should present a very distinct behavior from those of
infinite-range models. For the SG model on the pathological Bethe lattice
with finite connectivity, the controversy about the nature of the condensed
phase still persists. For instance, it was found that a replica-symmetric
solution is stable for zero field when {\em open} (uncorrelated) boundary
conditions are considered \cite{Thouless}, while the breaking of replica
symmetry is required to obtain a stable solution below $T_{c}$ when {\em %
closed }(correlated) boundary conditions are imposed to the system \cite
{Mottishaw}. Another line of approach in the study of short-range SG
behavior was developed after the work of Southern and Young \cite{Southern}
who succeeded to obtain, by using the Migdal-Kadanoff renormalization group
(MKRG) scheme, phase diagrams showing the presence of a SG phase in three
dimensions($d=3$), but not for $d=2$, indicating that the lower critical
dimension $d_l$ should lie in this interval. This latter approach, which can
be viewed as an approximation for real systems, was applied to investigate
the exponents required to describe the transition to the condensed phase. On
the other hand, based on the scaling theory, it was found that SG obeying
symmetric distributions is characterized by four independent exponents, the
thermal and the {\em chaotic} ones at $T=0$ and $T=T_{c}$ \cite{Hilhorst}.
Within this approach the {\em chaos} exponents govern the sign-changing of
the effective coupling of two spins, a distance $L$ apart, with the
temperature. The interpretation for this phenomenon of critical chaos was
done in the framework of the {\em droplet }theory\cite{Thill}.

Another point of view of the MKRG\ was explored since it was proved that its
renormalization-group equations are exact for the Ising model on a family of
diamond hierarchical lattices (DHL) \cite{Bleher}. The study of spin systems
on such exotic lattices, whose coordination number varies from two to
infinity, acquires relevance because exact solutions can be obtained and are
well-controlled. Motivated by this, we generalized the method proposed by
Morgado {\em et al. }\cite{Morgado} (MCC) to investigate the structure of
the local Edwards-Anderson (EA) order parameter of the Ising SG model on
these lattices. Through this method, we are able to calculate the local
magnetization at each site by an exact recursion procedure and furthermore,
the {\em local} EA order parameter associated to each site. The MCC method
was firstly applied to the pure ferromagnetic Ising model on a two-connected
DHL, revealing the multifractal structure of the local magnetization {\em at
the critical point} \cite{Morgado}. Moreover, they showed that there is an
infinite set of $\beta $ critical exponents for the local magnetization
which are linearly related to the $\alpha $-H\"older exponent of the $%
F(\alpha )$ multifractal spectrum. This work was further generalized for a $%
p $-connected DHL with $q$ intermediated sites within each connection \cite
{Osmundo}. The MCC method was also applied to investigate the multifractal
and critical properties of the q-state ferromagnetic Potts model on the $p$%
-connected DHL \cite{Ladario}.

In the present paper we extend the preliminary results obtained by Coutinho%
{\em \ et al.}, \cite{Coutinho} where the $F(\alpha )$-spectra of the SG
Ising model with an initial Gaussian distribution was studied for lattices
up to twelve generations. Now we consider four distinct initial
distributions of couplings ({\em Bimodal, Gaussian, Uniform and Exponential}%
), with lattices up to sixteen generations, investigating the temperature
dependence of the $F(\alpha )$-spectrum around $T_{c}$. In section II, the
model Hamiltonian is presented and the corresponding critical temperatures
associated with each distribution are obtained. In section III, we give full
details of the generalization of the MCC method for the SG case, as well as
about the numerical procedure used to obtain the EA order parameter profiles
for the four distinct initial distributions of couplings. In section IV, the
multifractal properties of the EA order parameter are obtained and the
corresponding $F(\alpha )$-functions calculated by distinct initial
distributions are compared. Furthermore, we investigate the temperature
behavior of the boundaries of the $F(\alpha )$-functions, above and below
the critical point, corresponding to each distribution. Finally, the
conclusions are summarized in section V.

\section{The Hamiltonian and the Migdal-Kadanoff
 renormalization-group procedure}

\begin{figure}
\leavevmode
\vbox{%
\epsfxsize=4cm
{\small \bf   
\unitlength=0.7 pt
\begin{picture}(200,200)(-90, -50)
\put(0,0){\circle{10}}
\put(110,0){\circle{10}}
\put(60,60){\circle*{10}}
\put(80,60){\circle*{10}}
\put(100,60){\circle*{10}}
\put(110,60){\circle*{2}}
\put(120,60){\circle*{2}}
\put(130,60){\circle*{2}}
\put(160,60){\circle*{10}}
\put(0,120){\circle{10}}
\put(110,120){\circle{10}}
\put(63.2,63.84){\line(5,6){43.6}}
\put(82.23,64.47){\line(1,2){25.54}}   
\put(100.82,64.93){\line(1,6){8.36}}   
\put(156.8,63.84){\line(-5,6){43.6}}   
\put(106.8,3.84){\line(-5,6){43.6}}   
\put(107.77,4.47){\line(-1,2){25.54}}   
\put(109.18,4.93){\line(-1,6){8.36}}   
\put(113.2,3.84){\line(5,6){43.6}}
\put(0,5){\line(0,120){110}}
\put(25,60){\makebox(0,0){$\longrightarrow$}}
\put(0,-15){\makebox(0,0){$\mu $}}
\put(0,135){\makebox(0,0){$\mu' $}}   
\put(60,48){\makebox(0,0){\scriptsize $\sigma_{1}$}}   
\put(78,48){\makebox(0,0){\scriptsize $\sigma_{2}$}} 
\put(94,48){\makebox(0,0){\scriptsize $\sigma_{3}$}}
\put(160,48){\makebox(0,0){\scriptsize $\sigma_{p}$}} 
\put(110,-15){\makebox(0,0){$\mu$}}
\put(110,135){\makebox(0,0){$\mu'$}}
\end{picture}

}}
\caption{General diamond hierarchical lattice basic unit, with $p$
parallel paths, each one composed by two bonds and one internal site
(scaling factor $2$).}
\label{Fig1}
\end{figure}
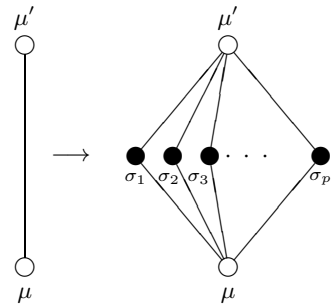

Let us consider the nearest-neighbor short-range Ising spin-glass model on a
general $p$-connection diamond hierarchical lattice. The hierarchical
lattices are connected graphs, recursively constructed by replacing all
bonds at each generation by a basic unit, the starting point being the basic
unit itself (first generation). The DHL basic unit is designed by two
root-sites coupled through $p$ connections {\em in parallel,} each of them
composed by two bonds {\em in series} via one internal site (scaling factor
2), as schematically shown in Fig.\ \ref{Fig1}. The {\em graph-fractal dimensions} of such lattices are given by $d=1+\ \ln p/\ln 2.$

The Hamiltonian of the present model is given by

\begin{equation}
\label{equation01}-\beta {\cal H}=\sum\limits_{<i,j>}K_N(i,j)\sigma
_i\sigma _j, 
\end{equation}

\noindent 
where $K_N(i,j)=-\beta J_N(i,j)$ is the reduced exchange coupling constant
between nearest neighbor spins of a $N$-generation DHL, $J_N(i,j)$ being the 
$N^{th}$-step Migdal-Kadanoff renormalized coupling constant obtained from
the original distribution. It is worth to mention that the real-space
Migdal-Kadanoff renormalization-group transformation on $d$-dimensional
Bravais lattices is known to be equivalent to the exact solution on the DHL
with a $d$-{\em graph fractal dimension} \cite{Bleher}. The MKRG
transformation of a $p$-connected DHL with general non-uniform coupling
constants yields an effective coupling given by \cite{Southern},

\begin{equation}
\label{equation02}K_N^{eff}(\mu ,\mu ^{\prime })=\frac
12\sum\limits_{l=1}^p\ln \left\{ \frac{\cosh \left[ K(\sigma _l,\mu
)+K(\sigma _l,\mu ^{\prime })\right] }{\cosh \left[ K(\sigma _l,\mu
)-K(\sigma _l,\mu ^{\prime })\right] }\right\} , 
\end{equation}

\noindent 
which can be rewritten in terms of more convenient variables as

\begin{equation}
\label{equation03}t_{\mu ,\mu ^{\prime }}=\tanh \left[ \sum\limits_{l=1}^p
tanh^{-1}\left( t_{\sigma ,\mu }t_{\sigma ,\mu ^{\prime }}\right)
\right] ,
\end{equation}

\noindent where $t_{x,y}=\tanh K_{x,y}$ is called thermal transmissivity.
In Eq.\ (\ref{equation02}) $\sigma _l$ labels the spins variables of the $p$
internal sites of a basic unit while $\mu $ and $\mu ^{\prime }$ denote the
root ones (see Fig.\ \ref{Fig1}). 

The critical temperature for the spin-glass model can be numerically
obtained either from eqs.(\ref{equation02}) or (\ref{equation03}), by
monitoring the width of the iterated distribution \cite{Southern}. For the
case of Eq.\ (\ref{equation03}), in the zeroth step a pool of $M$ random
initial transmissivities ($\{t_i\}=\tanh (\beta \{J_i\});$ $i=1,2,...,M)$ is
generated, with the coupling constants $\{J_i\}$ following a given
probability distribution. In the first step Eq.\ (\ref{equation03}) is
iterated $M$ times by choosing at random $2p$ initial transmissivities; the
new $M$-valued pool represents the renormalized thermal transmissivity
distribution. This process may be repeated and renormalized distributions
can be numerically followed by computing its moments at each step\cite
{Southern} and a flow diagram may be constructed, e.g., in the
transmissivity {\em versus} variance plane\cite{Curado}. Since the spin
glass {\em fixed-point distribution} is not analytically known, we
considered four initial symmetric distributions of interest, namely the {\em %
Gaussian}, {\em Bimodal}, {\em Exponential }and {\em Uniform} ones, defined
below,

\begin{mathletters}
\label{equation04}

\begin{equation}
\begin{array}[t]{lr}
P(J_{i,j})=\frac 1{\sqrt{2\pi }}\exp (-\frac 12\ J_{i,j}^2) &  \makebox{(Gaussian),}
\end{array}
\end{equation}

\begin{equation}
\begin{array}[t]{lr}
P(J_{i,j})=\frac 12\left[ \delta (J_{i,j}-1)+\delta (J_{i,j}+1)\right] &  
\makebox{(Bimodal),} 
\end{array}
\end{equation}

\begin{equation}
\begin{array}[t]{lll}
P(J_{i,j})=\frac 1{\sqrt{2}}\exp (-\sqrt{2}|J_{i,j}|) &  & \makebox{(Exponential),}
\end{array} 
\end{equation}
and
\begin{equation}
\begin{array}[t]{lll}
P(J_{i,j})=\left\{ 
\begin{array}{lllll}
\frac 1{2\sqrt{3}} &  &\makebox{if} &  &   -\sqrt{3}\leq J_{i,j}\leq \sqrt{3} \\
 0                 &  &             &  &   otherwise 
\end{array}
 \right\} &       &  \makebox{(Uniform)}.
\end{array} 
\end{equation}
\end{mathletters}

\noindent 
To each initial distribution a pseudo-critical temperature is associated,
for which the flow will converge to the critical point characterized by the
''fixed-point'' distribution which is numerically known \cite{Mckay}. For
temperatures very close but below (above) this pseudo-critical temperature
the flow will at first, approach the ''fixed point'' and then turn on to the
spin-glass (paramagnetic) fixed point characterized by infinite (zero)
variance and zero mean. In Table 1, we present our numerical estimates for
these pseudo-critical temperatures obtained from the initial distributions
given by eqs.(\ref{equation04}) in the case $d=3$ ($p=4$), compared to other
values reported in the literature.

\section{The local magnetization and  the local EA order parameter}

\subsection{The generalized method for spin glasses}

The aim of this method is to establish a recursion relation between the
value of the local magnetization of the internal site belonging to a
connection of a given basic unit of the DHL with the local magnetization of
its root sites. If this is achieved, one can consider a finite DHL with $N$
generations with Ising spin variables, described by the Hamiltonian given in
Eq.\ (\ref{equation01}), with nearest-neighbor random exchange coupling
constants chosen from a given initial distribution, and renormalize it $N-1$
times storing at each step all renormalized coupling constants. Then, taking
arbitrarily initial magnetizations (corresponding to the spin-glass boundary
conditions) for the root sites of the first generation, we can successively
calculate the local magnetizations of each site of the $N$-generation DHL,
at a given temperature and for a chosen initial coupling-constant
distribution.

\begin{figure}
\leavevmode
\vbox{%
\epsfxsize=4cm
{\small \bf   
\unitlength=0.7 pt
\begin{picture}(200,200)(-20, -50)
\put(110,0){\circle{10}}
\put(60,60){\circle{10}}
\put(80,60){\circle{10}}
\put(100,60){\circle*{2}}
\put(110,60){\circle*{2}}
\put(120,60){\circle*{10}}
\put(130,60){\circle*{2}}
\put(150,60){\circle*{2}}
\put(160,60){\circle{10}}
\put(110,120){\circle{10}}
\put(63.2,63.84){\line(5,6){43.6}}
\put(82.23,64.47){\line(1,2){25.54}}   
\put(119.18,64.93){\line(-1,6){8.36}}   
\put(156.8,63.84){\line(-5,6){43.6}}   
\put(106.8,3.84){\line(-5,6){43.6}}   
\put(107.77,4.47){\line(-1,2){25.54}}   
\put(110.82,4.93){\line(1,6){8.36}}   
\put(113.2,3.84){\line(5,6){43.6}}
\put(60,48){\makebox(0,0){\scriptsize $\sigma_{1}$}}   
\put(77,48){\makebox(0,0){\scriptsize $\sigma_{2}$}} 
\put(95,48){\makebox(0,0){\scriptsize $\sigma_{3}$}}
\put(160,48){\makebox(0,0){\scriptsize $\sigma_{p}$}} 
\put(95,0){\makebox(0,0){$\mu$}}
\put(95,120){\makebox(0,0){$\mu'$}}
\put(115,60){\oval(140,120)[r]}
\put(195,60){\makebox(0,0){$K'$}}
\put(110,140){\makebox(0,0){$\downarrow $}}
\put(125,138){\makebox(0,0){$h_{\mu'}$}}
\put(110,-20){\makebox(0,0){$\uparrow$}} 
\put(125,-22){\makebox(0,0){$h_{\mu}$}}           
\end{picture}
}}
\caption{Schematic representation of the equivalent system constructed
by retaining a given basic unit of the final generation,  with
coupling constants $K_i$ $(i=1,...,2p)$. $\mu ,\mu ^{\prime }$ are the root
spins of the basic unit, $h_\mu $, $h_{\mu ^{\prime }}$ and $K^{\prime }$
are, respectively, the effective fields and the effective coupling generated
by the whole lattice. }
\label{Fig2}
\end{figure}
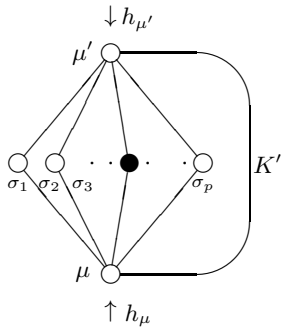

To obtain this recursion relation, let us consider a $N$-generation DHL and
look at an arbitrary basic unit introduced at the $N^{th}$-generation as
shown in Fig.\ \ref{Fig2}. This basic unit is connected to the lattice by its root sites ($\mu ,\mu ^{\prime }$). Therefore, the partition function of the
whole lattice can be written as

\begin{eqnarray}
\label{equation05}
\begin{array}{lll}

Z &=& Tr_{(\{\sigma _i\},\mu ,\mu ^{\prime })}^{}\exp \left\{ -\beta H^{\prime
}\right\}  \\ 
&=& Tr_{(\{\sigma _i\},\mu ,\mu ^{\prime })}^{}\exp \left\{ -\beta H\left(
\{\sigma _i\},\mu ,\mu ^{\prime }\right)\right\}\\
&&\times \exp \left\{ -\beta \left[ h_\mu \mu +h_{\mu
^{\prime }}\mu ^{\prime }+K^{\prime }\ \mu \mu ^{\prime}\right]\right\} 
\end{array}
\end{eqnarray} 
\noindent
 
where $\{\sigma _i\}$, $i=1,2...p$, denote the internal spins within each
connection, $\mu ,\mu ^{\prime }$ are the root spins of the basic unit, $%
h_\mu $, $h_{\mu ^{\prime }}$ and $K^{\prime }$ are, respectively, the
effective fields and the effective coupling acting upon the basic unit root
spins induced by the rest of the lattice. $H\left( \{\sigma _i\},\mu ,\mu
^{\prime }\right) $ is the internal Hamiltonian of the basic unit given by

\begin{equation}
\label{equation06}H\left( \{\sigma _i\},\mu ,\mu ^{\prime }\right)
=\sum\limits_{i=1}^p(K_i\mu +K_i^{\prime }\mu ^{\prime })\sigma _i, 
\end{equation} 

\noindent 
where $K_i,K_{\acute \imath }^{\prime }$ are the corresponding random
coupling constants between the $\sigma _i$'s and the $\mu ,$ $\mu ^{\prime }$
spins, respectively. The local magnetizations of all sites within a given basic unit can be easily evaluated for the effective model 
\begin{table}
\caption{Spin-glass pseudo-critical temperatures for different probabilities distributions within the Migdal-Kadanoff renormalization-group procedure, in $d=3$ spin glass.}
\label{Table01}
\begin{tabular}{|l|c|c|c|c|} 
 & { Gaussian} & {Bimodal} & {Exponential} & {Uniform} \\ \hline
This work  & 0.88  &  1.15  & 0.75  & 0.96  \\
Ref.\ \cite{Southern}   &  0.88$\pm$0.02 & 1.05$\pm$0.02 & - & 1.00$\pm$0.02\\
Ref.\ \cite{Curado} & 0.85  & 1.2 & 0.7 &  1.0  \\
Ref.\ \cite{Hilhorst} & 0.89 & -  & - & - \\
Ref.\ \cite{Campbell} &  0.83  &  1.15 &  0.71 & 0.94 \\ 
\end{tabular}
\end{table}

defined by the Hamiltonian $H^{\prime }$ by tracing over all spins
variables. However, since our main
concern is to establish a recursion relation between the internal-site
magnetization and these of the roots sites of a certain basic unit, the
procedure can be further simplified by focusing our attention to a single
connection and including the effects of the other connections onto the
effective fields and coupling, following the ideas of the decoration
transformation formalism\cite{fisher}. In this case, our system is
over-reduced to a single connection with an internal site under the action
of the effective fields and effective coupling induced by the remaining
lattice, as schematically shown in Fig.\ \ref{Fig2}.

For this over-reduced system, the Hamiltonian in defined by

\begin{equation}
\label{equation07}H^{\prime \prime }\left( \sigma ,\mu ,\mu ^{\prime
}\right) =(K_1\mu +K_2\mu ^{\prime })\sigma +h_\mu \mu +h_{\mu ^{\prime
}}\mu ^{\prime }+K^{\prime }\ \mu \mu ^{\prime }\,. 
\end{equation}

\noindent The magnetizations at each site are straightforwardly calculated,
giving

\begin{eqnarray}
\label{equation08}
<\mu >&=&Z^{-1}Tr\,\left[ \mu \exp (-\beta H^{\prime \prime
})\right] \nonumber \\
& & \nonumber \\
&=& \frac{(\tau _\mu +t\tau _{\mu ^{\prime }})+t_1t_2(\tau _{\mu
^{\prime }}+t\tau _\mu )}{1+t\tau _\mu \tau _{\mu ^{\prime }}+t_1t_2(t+\tau
_\mu \tau _{\mu ^{\prime }})},
\end{eqnarray}

\begin{eqnarray} 
\label{equation09}
<\mu ^{\prime }>&=&Z^{-1}Tr\,\left[ \mu ^{\prime }\exp
(-\beta H^{\prime \prime })\right] \nonumber \\
& & \nonumber \\
&=& \frac{(\tau _{\mu ^{\prime }}+t\tau
_\mu )+t_1t_2(\tau _\mu +t\tau _{\mu ^{\prime }})}{1+t\tau _\mu \tau _{\mu
^{\prime }}+t_1t_2(t+\tau _\mu \tau _{\mu ^{\prime }})}, 
\end{eqnarray}

\begin{eqnarray}
\label{equation10}
<\sigma >&=&Z^{-1}Tr\,\left[ \sigma \exp (-\beta H^{\prime
\prime })\right] \nonumber \\
 &  &  \nonumber \\
 &=&\frac{t_1(\tau _\mu +t\tau _{\mu ^{\prime }})+t_2(\tau
_{\mu ^{\prime }}+t\tau _\mu )}{1+t\tau _\mu \tau _{\mu ^{\prime
}}+t_1t_2(t+\tau _\mu \tau _{\mu ^{\prime }})}, 
\end{eqnarray}

\noindent where $t=\tanh (K^{^{\prime }})$, $t_i=\tanh (K_i)$ ($i=1,2)$ and $%
\tau _\alpha =\tanh (h_\alpha )$ for $\alpha =\mu ,\mu ^{\prime }$. Now,
from Eqs. (\ref{equation08}) and (\ref{equation09}) we can write the unknown
variables ($t,$ $\tau _\mu $ and $\tau _{\mu ^{\prime }}$) as a function of $%
<\mu >$ and $<\mu ^{\prime }>$ and $Z$, and substitute them in Eq.\ (\ref
{equation10}) to end up with the recursive equation

\begin{equation}
\label{equation11}<\sigma >=\frac{t_1(1-t_2^2)}{1-t_1^2t_2^2}<\mu >+\frac{%
t_2(1-t_1^2)}{1-t_1^2t_2^2}<\mu ^{\prime }>\,. 
\end{equation}

We emphasize that if the sites of a hierarchical lattice are properly
addressed, Eq.\ (\ref{equation11}) establishes a recursive equation between
the local magnetization of the sites belonging to the last generation and
the ones of previous generations. Moreover, its coefficients do not depend
on the unknown fields and couplings, but only upon the coupling constants
belonging to the chosen connection. This result is the main achievement of
this method.

\subsection{Numerical Procedure}

To analyze the structure of the local EA order parameter of our model, we
should calculate $<\sigma _i>^2$ for all sites and average them over many
samples, yielding to

\begin{equation}
\label{equation12}q_i^{EA}=[<\sigma _i>^2]_c\,, 
\end{equation}

\noindent where $[...]_c$ stands for the configurational average taken over
many independent initial distributions of couplings. To consider larger
lattices we have to go further in the renormalization steps. Since the
number of sites and bonds increases like $(2p)^N,$ the amount of computer
memory required to store the magnetizations and the coupling constants
during the intermediate steps will increase with such rate. In order to
maximize the number of renormalization steps, we look at the magnetization
structure of a subset of representative sites of the lattice. These sites
are the $2^N$ ones belonging to any shortest path connecting the roots
sites. The magnetization (and/or the EA local order parameter) structure of
this subset can be viewed as a representative {\em profile} of the whole
lattice. Since they are stochastically equivalent, we argue that by
averaging over many profiles, we should obtain the correct scenario for the
local EA order parameter of the considered model.

To calculate the profile of the local magnetization of a $N$ generation
hierarchical lattice we make use of Eq.\ (\ref{equation11}). To display the
profile, we have to label the sites of a given path, assigning the values of
its local magnetizations of the support set, due to the graph topological
nature of the hierarchical lattices. To proceed, we choose the set of site
labels by ($s,l$) belonging to the interval $[0,1]$, defined by $s\times
2^{-l}$, $s=1,3,5,...,(2^l-1)$, and $l$ labeling the generation ($%
l=1,2,...,N)$. For this choice the recursive equation can be written as

\begin{equation}
\label{equation13}<\sigma >_{s,l}=\Lambda _{s_1,l_1}<\mu >_{s_1,l_1}+\Lambda
_{s_j,l_j}<\mu ^{\prime }>_{s_j,l_j} 
\end{equation}

\noindent where $s_1=\frac 12(s\pm 1)$, $l_1=l-1$, $s_j=\frac 12(s\mp 1)$, $%
l_j=l-j$, with $j=2,3,...,l$. The coefficients of Eq.\ (\ref{equation13}) are
given by
\begin{mathletters}
\label{equation14}
\begin{equation}
\Lambda _{s_1,l_1}\ = 
\frac{t_{s,s_1}(1-t_{s,s_j}^2)}{1-t_{s,s_1}^2t_{s,s_j}^2}
\end{equation}
\begin{equation} 
\Lambda _{_{s_j,l_j}}\ =\frac{t_{s,s_j}(1-t_{_{_{s,s_1}}}^2)}{%
1-t_{_{}s,s_1}^2t_{s,s_j}^2\ } 
\end{equation}
\end{mathletters}

\noindent where $t_{s,s_j}=\tanh [K_l(s,s_j)]$, $K_l(s,s_j)$ being the
coupling constant between the spins at the positions $s.2^{-l}$ and the one
at $s_j.2^{-l_j}$.

To calculate an EA order parameter profile we must first generate the
coupling constants for each level. Due to the disordered nature of the
profile we make use of an equivalent stochastic procedure in order to save
computer memory at intermediate steps of the calculation. For a fixed
value of the temperature we create an initial distribution for the thermal
transmissivities, represented by a pool of $M$ random numbers ($M\cong
10\cdot 2^N$). At the $Nth$ level we choose randomly, from the initial
distribution, a set of $2^N$ couplings, which are stored to be used later in
the calculation of the site magnetizations.\ At the next level [$(N-1)th$
level], we obtain a renormalized distribution (new $M$ random numbers),
generated according to the renormalization Eq.\ (\ref{equation03}); from this
distribution, we pick randomly $2^{N-1}$couplings, which are also stored.
This process is carried for $N-1$ times, such that at the last level, only
two couplings are stored. Now, we make use of Eq.\ (\ref{equation13}), fixing
the initial values for the magnetization of the roots (zeroth generation)
and calculate the local magnetization of each level, using for the coupling
constants the values previously stored.

\subsection{The EA order parameter profiles}

\begin{figure}
\leavevmode
\vbox{%
\epsfxsize=8cm
\epsffile{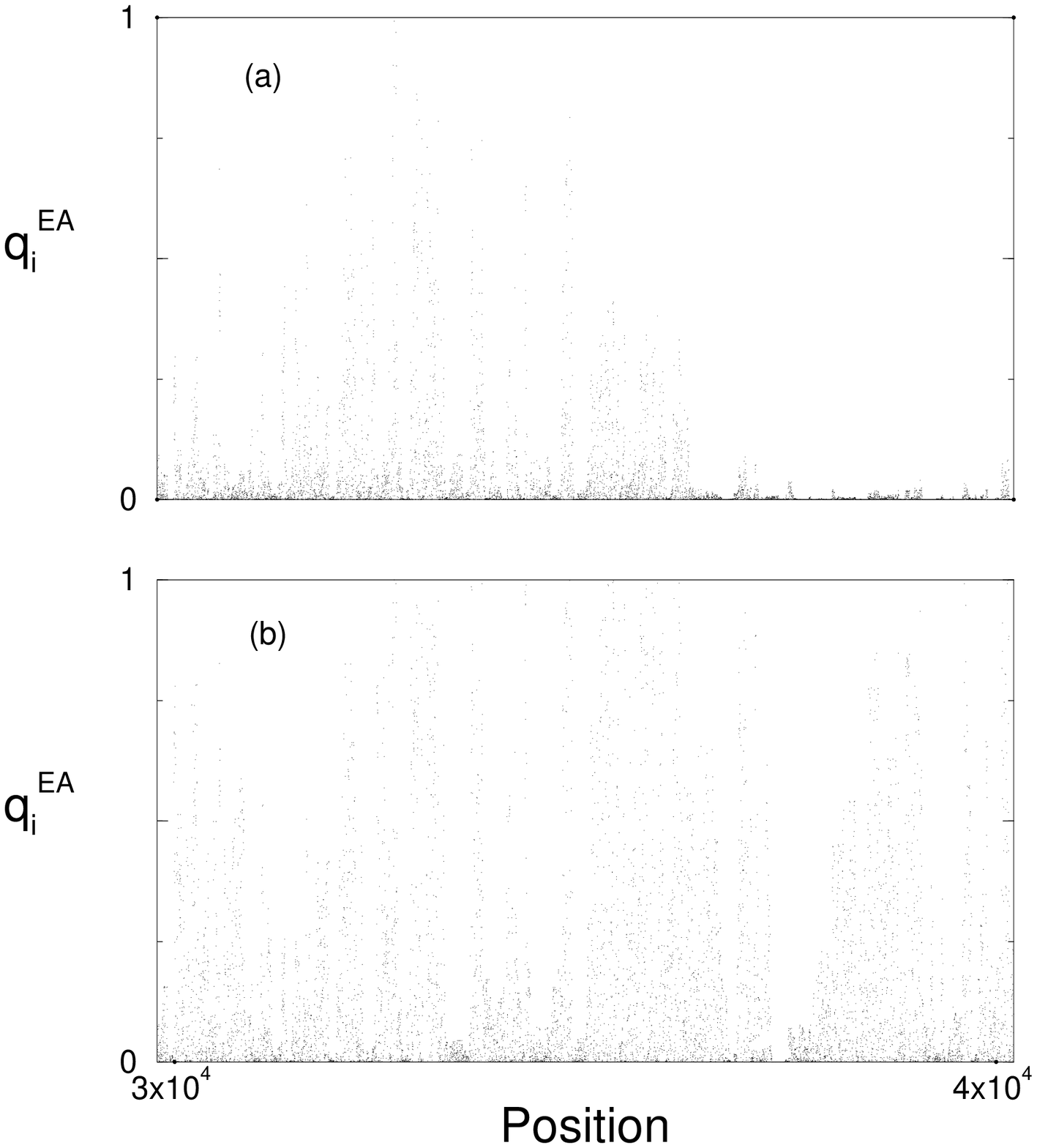}}
\caption{Portion of the profile of local EA order parameter for one
sample and a lattice with $N=16$ generations; the sites chosen belong to a
sub-set with positions in the range $i=3.0\times 10^4,...,4.0\times 10^4,$
selected from the set $i=1,2,...,2^N,$ corresponding to a given shortest
path connecting the two roots sites. (a) $T=T_{c}$ . (b) $T=0.9T_{c}$.}
\label{Fig3}
\end{figure}

\begin{figure}
\leavevmode
\vbox{%
\epsfxsize=8cm
\epsffile{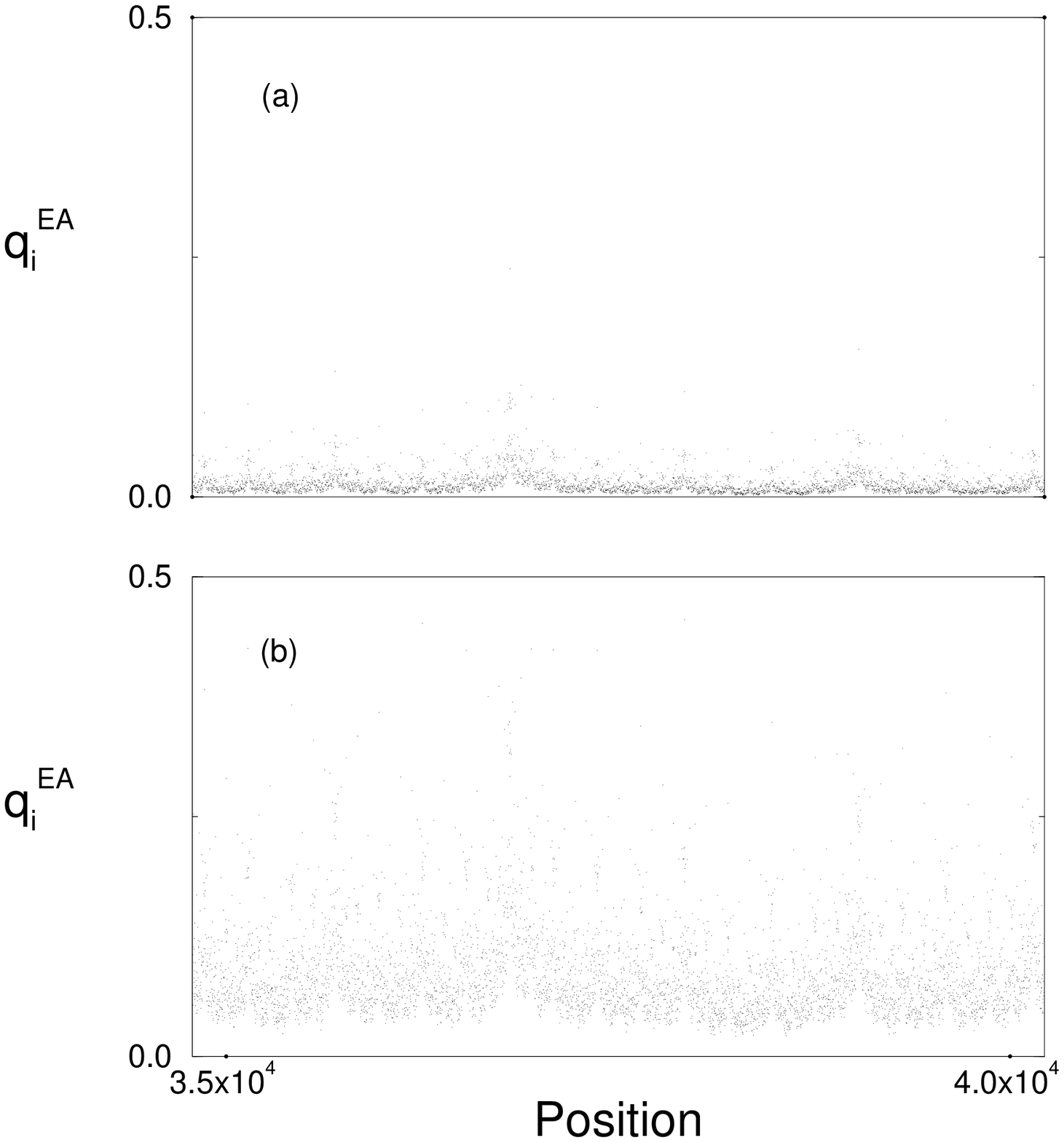}}
\caption{Portion of the profile of the average local EA order
parameter ($200$ samples) for a lattice with $N=16$ generations; the sites
chosen are the same as in Fig.\ \protect\ref{Fig3}. (a) $T=T_{c}$ . (b) $T=0.9T_{c}$.}
\label{Fig4}
\end{figure}

For each of the initial distributions defined in Eq.\ (\ref{equation04}), we
generated profiles at the corresponding critical temperature $T_{c}$ (see
table \ref{Table01}), as well as $T_1=0.9T_{c}$, $T_2=0.8T_{c}$ and $T_3=0.7T_{c}.$%
\ This was done for lattices with $d=3$ and $N=8$ up to $16$ hierarchies. In
Fig.\ \ref{Fig3} we display these profiles for the Gaussian distribution at the
temperatures $T_{c}$ and $T_1$ ($N=16$) for just one sample, whereas in Fig.\ \ref{Fig4}, the same is done for 200 samples. It is clear from these figures that the
disordered structure of the local EA\ order parameter of one sample
increases as we go further in the condensed phase. However, when the
configurational average is taken, the profile presents uniformities
reminiscent of the graph lattice symmetry, similar to what happens for the
pure model \cite{Morgado}. It is also evident from Fig.\ \ref{Fig4}, the increasing of 
$q_{}^{EA}=2^{-N}\sum_iq_i^{EA}$ , the mean value per site, as we decrease
the temperature. For all other distributions listed above ({\em Bimodal,
Exponential }and {\em Uniform}), the same qualitative behavior was observed 
\cite{Edvaldo}.

\section{\ The Multifractal Properties}

The high degree of discontinuity shown in the profiles suggests us the
multifractal analysis as a tool to investigate the singularities of the
measure constructed from the EA local order parameter, following the same
approach used for pure models\cite{Morgado,Osmundo,Ladario}. To obtain the
multifractal spectra ({\em F(}$\alpha ${\em )}-function), we first define a
measure by the normalized local EA\ order parameter

\begin{equation}
\label{equation15}\zeta _i^{EA}=\ \frac{q_i^{EA}}{\sum_iq_i^{EA}}\,, 
\end{equation}

\noindent and construct a parametrized family of normalized measures defined
by

\begin{equation}
\label{equation16}\mu _i^{EA_{}}=\frac{(\zeta _i^{EA})^q}{\sum_i(\zeta
_i^{EA})^q}\,. 
\end{equation}

The {\em F(}$\alpha ${\em )}-function{\em \ }is now obtained following the
method due to Chhabra and Jensen\cite{Chhabra}, where the {\em spectrum} is
obtained by varying the parameter $q$ and calculating

\begin{equation}
\label{equation17}F(\alpha _q)=\lim _{N\rightarrow \infty }\left\{ \frac{-1%
}{N\ln 2}\sum\limits_i\mu _i^{EA_{}}\ln \mu _i^{EA_{}}\right\} \,, 
\end{equation}

\begin{equation}
\label{equation18}\alpha _q=\lim _{N\rightarrow \infty }\left\{ \frac{-1}{%
N\ln 2}\sum\limits_i\mu _i^{EA_{}}\ln \zeta _i^{EA}\right\} \,. 
\end{equation}

In Fig.\ \ref{Fig5}, we display the corresponding {\em F(}$\alpha ${\em )}-functions%
{\em \ }for the profiles of one sample obtained from the initial
distributions listed in Eq.\ (\ref{equation04}), for temperatures at and
below $T_{c}$. We notice light variations of the spectra by changing the
temperature for the Gaussian, Bimodal and Uniform cases, whereas more
pronounced changes are observed for the Exponential case. In Fig.\ \ref{Fig6}, the
corresponding {\em F(}$\alpha ${\em )}-functions{\em \ }averaged over $200$
samples are shown for the Gaussian and Bimodal distributions at $T_{c}$ and $%
T_1=0.9T_{c}$. Minor changes are observed when we compare with Fig.\ \ref{Fig5}.

\begin{figure}
\leavevmode
\vbox{%
\epsfxsize=8cm
\epsffile{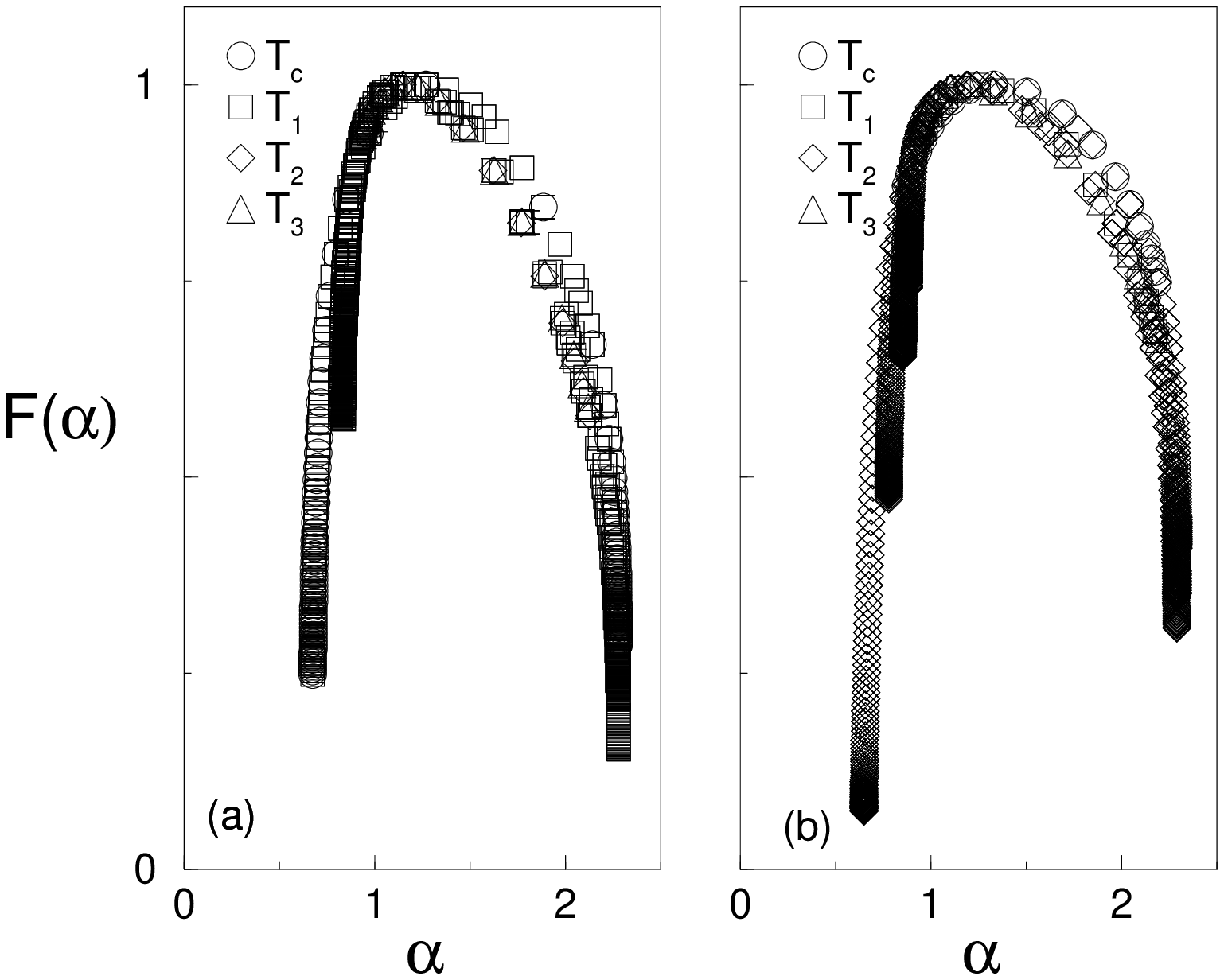}}
\end{figure}

\begin{figure}
\leavevmode
\vbox{%
\epsfxsize=8cm
\epsffile{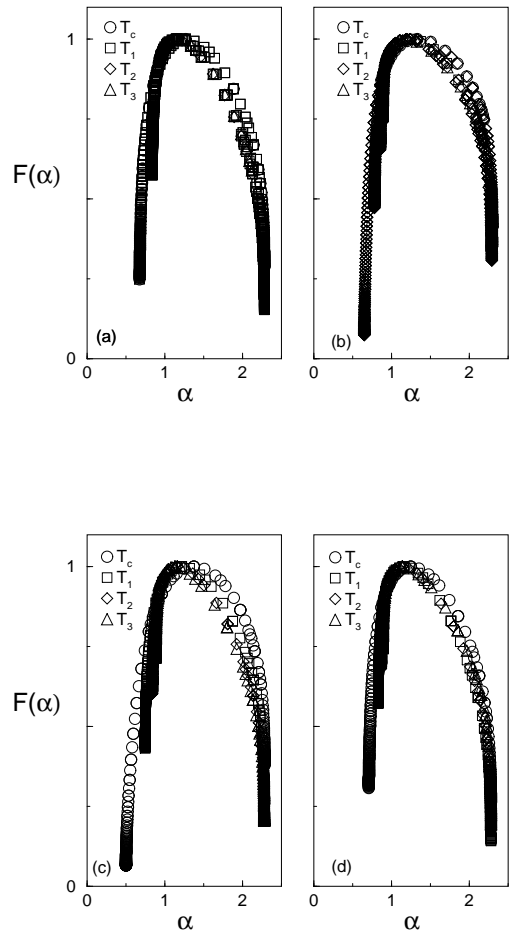}}
\caption{$F(\alpha )$-functions of the local EA order parameter
profile with one sample, for decreasing temperatures and different initial
distributions of coupling constants. (a) {\em Gaussian}, (b) {\em Bimodal,}
(c) {\em Exponential} and (d) {\em Uniform} distributions ($\bigcirc $ $T=T_{c}
$, $\Box $ $T_1=0.9T_{c}$, $\Diamond $ $T_2=0.8T_{c}$ and $\triangle $ $%
T_3=0.7T_{c}$).}
\label{Fig5}
\end{figure}

In Fig.\ \ref{Fig7}, we show in the same plot the {\em F(}$\alpha ${\em )}-functions at and
below $T_{c}$ for the four considered distributions. It is worth to call the
reader's attention to the universal character of the {\em F(}$\alpha ${\em )}%
-function within small deviations. Here we remark that although we have used
the renormalized distributions of couplings at each step of the calculation,
the influence of the initial distribution should be relevant, since in the
thermodynamic limit ($N\rightarrow \infty $), {\em half }of number of sites
in the profile belong to generation $N$ and their magnetizations are
calculated with the coupling constants introduced by the initial
distribution. For the whole lattice this influence should be even more
relevant; in this case  $\frac 78$ of the total number of sites
belong to generation $N$.

\begin{figure}
\leavevmode
\vbox{%
\epsfxsize=8cm
\epsffile{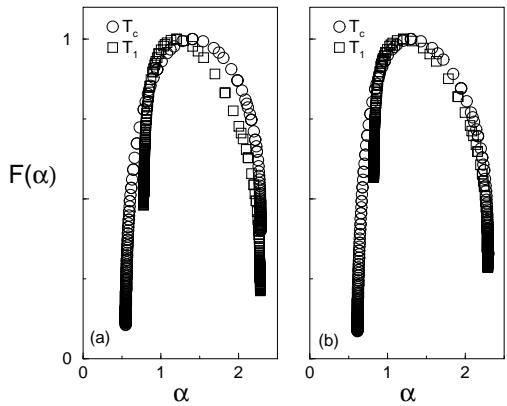}}
\caption{$F(\alpha )$-functions of the average local EA order
parameter profile (200 samples), at and below the critical temperature, for
the (a) {\em Gaussian} and (b) {\em Bimodal} distributions of coupling
constants ($\bigcirc $ $T=T_{c}$, $\Box $ $T_1=0.9T_{c}$).}
\label{Fig6}
\end{figure}

\begin{figure}
\leavevmode
\vbox{%
\epsfxsize=8cm
\epsffile{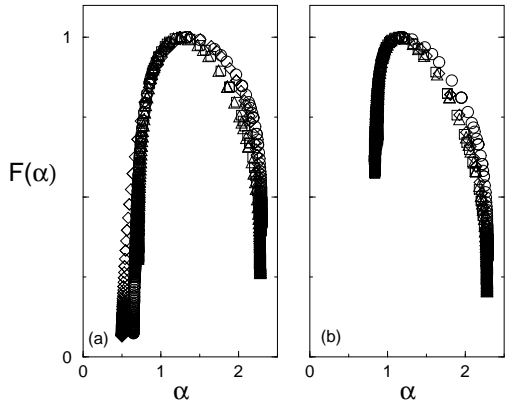}}
\caption{Comparison between the $F(\alpha )$-functions of the
local EA order parameter profiles at and below the critical temperature for
the $\bigcirc $ {\em Gaussian}, $\Box $ {\em Bimodal,} $\Diamond $ {\em %
Exponential} and $\triangle $ {\em Uniform} initial distributions of
coupling constants. (a) $T=T_{c}$ and (b) $T_1=0.9T_{c}$ .}
\label{Fig7}
\end{figure}

\begin{figure}
\leavevmode
\vbox{%
\epsfxsize=8cm
\epsffile{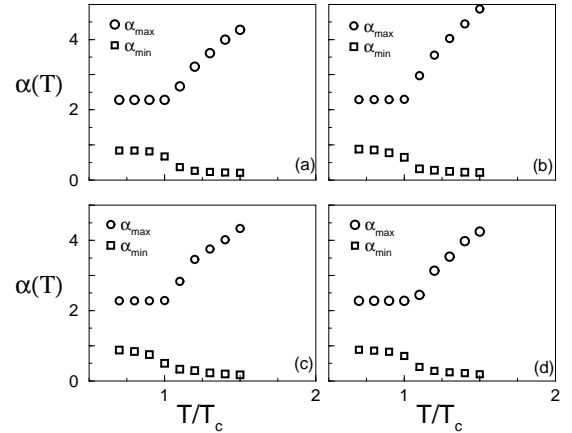}}
\caption{Upper ($\bigcirc $ $\alpha _{\max }$) and lower ($\Box $ $%
\alpha _{\min })$ bounds of the $F(\alpha )$ spectrum as a function of the
temperature for different distributions of coupling constants. (a) {\em %
Gaussian}, (b) {\em Bimodal,} (c) {\em Exponential} and (d) {\em Uniform}
distributions.}
\label{Fig8}
\end{figure}

In order to investigate the behavior of the {\em %
F(}$\alpha ${\em )} spectrum around $T_{c}$, we plot in Fig.\ \ref{Fig8} the dependence
of its upper and lower bounds as a function of the temperature, for all the
considered initial distributions. We notice that below $T_{c}$ a constant
value is observed for the upper bound ($\alpha _{\max }$), while small
deviations occur for the lower bound ($\alpha _{\min }$) close to $T_{c}$.
Nevertheless, for temperatures slightly above $T_{c}$ an abrupt increase is
observed for the upper bound while a small decrease occurs for the lower
bound, signalizing the SG transition. At $T\geq T_{c}$ the magnetization at
the majority of sites vanishes, being eliminated from the calculation of the
spectrum. Therefore, as we are dealing with a finite lattice one should
expect a finite but higher $\alpha _{\max }$, which is the exponent
governing the singularities of the set of smallest measures still present.
This is evidenced by the rapid increasing of $\alpha _{\max }$ with the
temperature, for $T>T_{c}$. On the other hand, the $\alpha _{\min }$ exponent
which governs the set of higher measures should remain finite to describe
the singularities of the measures belonging to the sites ''close'' to the
root sites (or{\em \ surface }sites). Those sites are the ones whose
magnetizations were calculated with at least one of the values imposed as
initial boundary conditions. The intermediate points of the spectra should
be {\em spurious} points, since the present algorithm\cite{Chhabra} used to
calculate the $F(\alpha )$-function is based on the method of moments and
tend to produce a top envelop of the actual spectrum\cite{Daniele}. In the
thermodynamic limit ($N\rightarrow \infty $), we expect to have no spectrum
except a point ($0,0$) corresponding to the non-vanishing values introduced
by the imposed boundary conditions.

\section{Conclusions}

We generalized the exact recursion method developed by Morgado {\em et al. }%
\cite{Morgado} and applied it to investigate the structure of singularities
of the local Edwards-Anderson order parameter of the short-range Ising
spin-glass model on diamond hierarchical lattices. Within this procedure,
the distribution of coupling constants is renormalized by the
Migdal-Kadanoff renormalization equation, which is exact in such classes of
lattices. The EA order parameter profiles ($2^N$ sites) for lattices up to $%
N=16$ hierarchies were calculated by considering four types of initial
distributions (Gaussian, Bimodal, Exponential and Uniform) and for
temperatures around the critical point. For $T\leq T_{c}$ the profiles of the
spin-glass condensed phase show a high degree of disorder and singularities
increasing for lower temperatures. The multifractal analysis was applied to
these profiles, revealing a large spectrum of exponents for the
singularities of the measure defined by the normalized local EA order
parameter. For each considered distribution of couplings, the $F(\alpha )$%
-function of the profiles show slight variations with the temperature in the
studied range of $0.7T_{c}\sim T_{c}$, with larger deviations occurring when the
temperature gets closer to $T_{c}$. These profiles reveal a high degree of
local disorder of the spin-glass condensed phase, a scenario not observed in
the pure case \cite{Morgado,Osmundo} for the same classes of lattices.
Moreover, small deviations are observed when the multifractal spectrum
obtained from distinct initial distributions of coupling constants are
compared at the same temperature. This suggests an {\em universal}
multifractal behavior of the present model for distinct initial
distributions of couplings, taking into account that $\frac 78$ of the total
number of sites belong to the last generation, whose local magnetizations
are calculated with the coupling constants introduced by the
not-yet-renormalized (initial) distribution. We have also studied the
temperature dependence of the range of the multifractal spectrum close to
the critical point. For $T\leq T_{c}$ the range of the $\alpha $-H\"older
exponent remains quite constant. However, when the temperature is higher
than $T_{c}$, one observes an abrupt change in the multifractal spectrum
signalizing the transition.

Contrary to the pure case, where a non-trivial multifractal behavior is
observed only at the critical temperature\cite{Morgado,Osmundo}, the
persistence of the $F(\alpha )$-function throughout the spin-glass phase
indicates the highly non-trivial character of such phase. Although we are
not able to associate the persistence of multifractality with any prediction
from the available theories to describe short-range spin glasses, the
present work evidences the contrast between the spin-glass and ferromagnetic
states and the critical nature of the spin-glass condensed phase.

\acknowledgments{
This research was supported by the CNPq, FINEP and CAPES ( Brazilian
governmental granting agencies). One of us (E. N. Jr.) is also grateful to
FACEPE (Pernambuco state granting agency) for the financial support under
the grant BFD-0505-1.05/95.}

\end{document}